# Saha Equation Normalized to Total Atomic Number Density

John W. Fowler[1]

Infrared Processing and Analysis Center
California Institute of Technology
1200 E. California Blvd., Pasadena, CA 91125, USA

**Abstract**

The Saha equation describes the relative number density of consecutive ionization levels of a given atomic species under conditions of thermodynamic equilibrium in an ionized gas. Because the number density in the denominator may be very small, special steps must be taken to ensure numerical stability. In this paper we recast the equation into a form in which each ionization fraction is normalized by the total number density of the atomic species, analogous to the Boltzmann equation describing the distribution of excitation states for a given ion.

Numerous astrophysical problems require computing the distribution of ionization states in ionized gaseous media (e.g., stellar atmospheres and interiors). In many cases, local thermodynamic equilibrium may be assumed or at least used as an approximation from which to add perturbations due to departures from equilibrium. The equation describing this distribution for thermodynamic equilibrium was originally formulated by M. Saha (1921). A rigorous derivation using statistical mechanics was provided by R.H. Fowler (1923), and various other approaches have been used (e.g., D.H. Menzel, 1935, quoted by L.H. Aller, 1963). We will begin with the equation as it is typically encountered for a single atomic species:

$$\frac{N_{i+1}}{N_i} = \frac{(2\pi m_e kT)^{\frac{3}{2}}}{N_e h^3} \frac{2u_{i+1}}{u_i} e^{\frac{-\chi_{i+1}}{kT}}$$

where $N_i$ is the number density of ions in ionization state $i$ (neutral state = 0), $N_e$ is the number density of electrons, $m_e$ is the electron mass, $k$ is the Boltzmann constant, $T$ is the temperature, $h$ is Planck's constant, $u_i$ is the partition function for state $i$, and $\chi_i$ is the ionization potential, the energy required to ionize the atom from state $i$-1 to state $i$. If the ideal gas law is an acceptable approximation (e.g., as it usually is for stellar atmospheres), then it is often more convenient to use the electron pressure $P_e = N_e kT$, and to reduce notational clutter, we define a "Saha constant" $C_s \equiv 2k(2\pi m_e k)^{3/2}/h^3$ to obtain

$$\frac{N_{i+1}}{N_i} = \frac{C_s T^{\frac{5}{2}}}{P_e} \frac{u_{i+1}}{u_i} e^{\frac{-\chi_{i+1}}{kT}} \quad \text{or} \quad \frac{C_s T^{\frac{3}{2}}}{N_e k} \frac{u_{i+1}}{u_i} e^{\frac{-\chi_{i+1}}{kT}}$$

---

[1] email address: jwf@ipac.caltech.edu

Furthermore, the first factor on the right-hand side does not depend on ionization level, so the notation can be simplified further by defining $\gamma = \gamma(T,P_e) \equiv C_s T^{5/2}/P_e$ or $\gamma(T,N_e) = C_s T^{3/2}/(N_e k)$:

$$\frac{N_{i+1}}{N_i} = \gamma \frac{u_{i+1}}{u_i} e^{\frac{-\chi_{i+1}}{kT}}$$

The fact that the number density ratio applies only to consecutive ionization states is inconvenient, since one must be sure that $N_i > 0$ before attempting to compute the ratio. Techniques to handle this have been developed (e.g., the "dominant potential method"), but many applications require knowledge of the distribution over all ionization states, and computing these ionization fractions numerically from a set of ratios after navigating around possible divisions by extremely small numbers is less efficient computationally than computing them from a single formula normalized to the total atomic number density.

This can be done if the Saha equation is recast into the appropriate form. The derivation of this form was published in 1970 by the author in an admittedly obscure journal intended primarily for graduate students to practice writing scientific articles (Fowler, 1970). The Saha equation in this new form was used in the author's doctoral dissertation, but the derivation was not included in the thesis or subsequent related publications (e.g., Fowler, 1974), and although some reprints of the 1970 article were sent out as a result of several requests, the equation never achieved traction as far as the author has been able to determine with a reasonable literature search. If other authors have independently produced equivalent derivations, then the author apologizes for overlooking them, but clearly our combined efforts are needed to get the improved form into the mainstream.

We begin by observing that

$$\frac{N_{i+2}}{N_i} = \frac{N_{i+1}}{N_i} \frac{N_{i+2}}{N_{i+1}}$$

and in general,

$$\frac{N_{i+n}}{N_i} = \prod_{m=1}^{n} \frac{N_{i+m}}{N_{i+m-1}}$$

Thus the ratio on the left-hand side is the product of $n$ traditional Saha equations. This product will have the following properties: (a.) since each ratio contributes a factor of $\gamma$, the product will contain $\gamma^n$; (b.) each ratio contributes a ratio of partition functions, all of which cancel except those with the lowest and highest index, leaving a factor of $u_{i+n}/u_i$; (c.) the product of all the exponentiated ionization potentials accumulates a sum of those potential energies in the argument of the exponential. Defining

$$\beta_i \equiv \sum_{j=0}^{i} \chi_j, \quad \chi_0 \equiv 0$$

this factor is $\exp((\beta_i - \beta_{i+n})/kT)$. Therefore the ionization equation for nonconsecutive states becomes

$$\frac{N_{i+n}}{N_i} = \gamma^n \frac{u_{i+n}}{u_i} e^{\frac{\beta_i - \beta_{i+n}}{kT}}$$

Now let us use $m \equiv i + n$:

$$\frac{N_m}{N_i} = \gamma^{m-i} \frac{u_m}{u_i} e^{\frac{\beta_i - \beta_m}{kT}}$$

This form can be useful if one has found the dominant potential and defined $i$ to be the corresponding state, since $m$ need not be greater than $i$, and the fractions can all be computed relative to the dominant level. By observing that this equation separates with respect to the ionization index, however, we can write it

$$\frac{N_m}{\gamma^m u_m e^{-\frac{\beta_m}{kT}}} = \frac{N_i}{\gamma^i u_i e^{-\frac{\beta_i}{kT}}}$$

hence each side must be equal to a separation constant (i.e., not dependent on ionization index):

$$\frac{N_i}{\gamma^i u_i e^{-\frac{\beta_i}{kT}}} = \lambda(T, P_e \text{ or } N_e)$$

or

$$N_i = \lambda \gamma^i u_i e^{-\frac{\beta_i}{kT}}$$

Defining

$$G_i(T, P_e \text{ or } N_e) \equiv \gamma^i u_i e^{-\frac{\beta_i}{kT}}$$

then with

$$G = \sum_i G_i$$

$$N = \sum_i N_i = \lambda \sum_i G_i = \lambda G$$

where the summations are from zero to the atomic number, we have

$$\frac{N_i}{N} = \frac{\gamma^i u_i e^{-\frac{\beta_i}{kT}}}{G}$$

i.e., the separation constant $\lambda$ has canceled out, and we have an equation with the same form as Boltzmann's distribution of excitation states for a given ion, but in this case, it is the distribution of ionization states for a given atomic species.

The author would like to thank Frank Masci for his assistance with this manuscript.

**References**

Aller, L.H., *The Atmospheres of the Sun and Stars*, Chapter 3, The Ronald Press Company, 1963

Fowler, J.W., *Ionization Equilibrium in Stellar Plasmas*, Maryland Astronomical Journal Vol. 1 No.3, 42, 1970.

Fowler, J.W., *A Line-Blanketed Model Stellar Atmosphere of Sirius*, ApJ **188**, 295-307, 1974

Fowler, R.H., *Phil. Mag.*, **45**, 1, 1923

Menzel, D.H., and Pekeris, C.L., *M.N.*, **96**, 77, 1935

Saha, M., *Phil. Mag.*, **41**, 267, 1921